# Ab initio investigation of lattice dynamics of fluoride scheelite LiYF$_4$

Benoit Minisini, Qiuping A. Wang and François Tsobnang.

Institut Supérieur des Matériaux et Mécaniques Avancées du Mans, 44 Av. Bartholdi, 72000 Le Mans, France

**Abstract:**

We report on the phonon dynamics of LiYF$_4$ obtained by direct method using first principle calculations. The agreement between experimental and calculated modes is satisfactory. An inversion between two Raman active modes is noticed compared to inelastic neutron scattering and Raman measurements. The atomic displacements corresponding to these modes are discussed. Multiple inversions between Raman and infrared active groups are present above 360 cm$^{-1}$. The total and partial phonon density of state is also calculated and analyzed.

**I Introduction**

Researches on YLiF$_4$ crystals are strongly linked to laser technology. The first structural data obtained by Thomas et al. date from 1961[1], just one year after the demonstration of the first laser. At ambient pressure, the crystal cell of YLiF$_4$ is tetragonal with space group I4$_1$/a (C$_{4h}^6$). This phase is commonly named the scheelite structure in reference of the CaWO$_4$ crystal. Lithium ions (Li$^+$) are in the center of tetrahedrons composed by 4 fluoride ions (F$^-$). Yttrium ions (Y$^{3+}$) are in the center of polyhedrons composed by 8 F$^-$. Y$^{3+}$ can be substituted by rare earth presenting an oxidation state of +3, such as Erbium (Er$^{+3}$) [2] or Thulium (Tm$^{+3}$) [3], providing good matrix for upconversion laser. The efficiency of this kind of laser relies on intraionic and interionic process of relaxation that strongly depends on the host matrix[4]. This relationship is evidenced particularly by the multiphonon relaxation process implying electron-phonon coupling[5]. Consequently, a fair knowledge of the structural and dynamics properties of the host matrix is crucial for the development of host matrix.

To this end many studies have been carried out on the subject. Phonon frequencies were measured by Raman and IR spectra [6][7][8][9][10]. These methods give information at the center of the Brillouin zone. Inelastic neutron scattering measurement is needed to obtain the complete phonon dispersion curves that are essential to a good understanding of the global vibrational and relevant properties. This was done for $LiYF_4$ by Salaün et al.[11].

Besides experimental work, numerical methods have been developed. Among them we can notice empirical methods, such as rigid ion model (RIM). Using this method, Salaün et al[11] and Sen et al.[12] performed lattice dynamical calculations on $LiYF_4$ providing a large number of interesting results about lattice vibration. Obviously, the correctness and precision of this model is limited by the empirical parameters.

Density functional theory is an empirical free parameter methods whose usefulness and predictive ability in different fields[13][14] are known since a long time. Recently, the association of DFT with different techniques such as linear response method[15][16] or direct methods[17][18][19] allows to evaluate phonon dispersion curves without empirical parameter. In particular Parlinsky et al.[20][21] developed a direct method where the force constant matrices are calculated via the Hellmann-Feynman theorem in total energy calculations.

In this work we present a first principle investigation of $YLiF_4$ in its scheelite phase. DFT associated with projector augmented wave (PAW) and direct method were used. Cell parameters, phonon dispersion curve, phonon density of state are discussed and compared with previous experimental or numerical results. To our knowledge, this is the first ab initio calculation of $LiYF_4$ lattice dynamics.

**II Methodology** :

Cell parameter and atomic positions of the initial structure were obtained from experimental results by E. Garcia and R.R Ryan[22]. All calculations were carried out with the VASP[23] code, based on DFT [24][25], as implemented within MEDEA[26] interface. Here the generalized gradient approximation (GGA) through the Perdew Wang 91 (PW91)[27] functional and projector augmented wave (PAW)[28] were employed for all calculations. Electronic occupancies were determined according to a Methfessel-Paxton scheme[29] with an energy smearing of 0.2 eV.

The crystal structure was optimized without the constraints of the space group symmetry at 0 Gpa until the maximum force acting on each atom dropped below 0.002 eV/Å. The self consistent field (SCF) convergence criterion was set at $10^{-6}$ eV. High precision calculations, as defined in VASP terminology, were performed with a basis set of plane wave truncated at a kinetic energy of 700eV. The Pulay stress[30] obtained on the unit cell was -4 MPa and the convergence of the total energy was within 0.4 meV/atom compared to an energy of 750 eV. Brillouin zone integrations were performed by using a 3X3X3 k-points Monkorst-Pack[31] grid leading to a convergence of the total energy within 0.1 meV/cell compared to a 7x7x3 k-point mesh.

PHONON code[19], based on the harmonic approximation, as implemented within MEDEA[26] was used to calculate the phonon dispersion. From the optimized crystal structure, a 2X2X1 supercell consisting of 96 atoms, was generated from the conventional cell to account for an interaction range of about 10 Å. The asymmetric atoms were displaced by +/- 0.03 Å leading to 14 new structures. The dynamical matrix was obtained from the forces calculated via the Hellmann-Feynman theorem. **G** point and medium precision, as defined in VASP terminology, were used for theses calculations. The error on the force can perturb the translation-rotational invariance condition. Consequently, this condition has to be enforced. A strength of enforcement of the translational invariance condition was fixed at 0.1 during the derivation of all force constants. The longitudinal optical (LO) and transversal optical mode (TO) splitting was not investigated in this work. Consequently, only TO modes at the **Γ** point were obtained.

**III Results and discussion:**

**I.a Structural parameters.**

Table 1 shows calculated and previous experimental or numerical structural properties of LiYF$_4$. Compared to the most recent experimental data[32][33] our calculated volume is over-estimated. Nevertheless, the *c/a* axial ratio, whose evolution is significant in pressure induced transition phase, is close to experimental results.

DFT results are strongly dependent on the approximation of the exchange correlation term. It's known that local density approximation (LDA) favorizes high electron densities resulting in short bonds prediction and so in low equilibrium volume. Results obtained by Li et al.[35] and Ching et al. [36] owing to LDA illustrate this behavior. GGA corrects and sometimes over-corrects the failures of the LDA. That's why cell parameters obtained using PW91 differ from experimental results. At least two reasons explain why our results are at variance with Li et al.[35]. The first one is due to the utilization of different parameters such as the energy cut off. The other one can be attributed to the difference of method to evaluate the equilibrium volume. Indeed, during a structure optimization the convergence criterion is set on the stress. Consequently, the lattice constants are obtained from a cell at its minimum of stress. But a minimum of stress does not mean a minimum of energy. Lattice constants from a cell at its minimum of energy, as performed by Li et al.[35], are derived through fitting of an energy versus volume curve. Conventional cell contains 16 fluorine (18.99 g/mol), 4 Lithium (7.01 g/mol) and 4 Yttrium (88.90 g/mol). Consequently we obtained a density of 3.86 Mg/m$^3$ which is about 3% lower than the density evaluated from cell parameters given by X ray. We can notice that Blanchfield and Saunders[37] measured a floatation density of 3.98 Mg/m$^3$. Bond lengths are presented in Table 2. As a result of GGA over correction, they are 3% longer than experimental distance.

**III.b Lattice dynamic.**

The phonon dispersion curves along several lines of high symmetry for LiYF$_4$ structure at zero pressure are shown in Figure 1. To evaluate our calculated phonon dispersion curves, the acoustic branches will be first compared to results extracted from ultrasonic measurements and rigid ion models (RIM). Then the modes at the **G** point will be compared to experimental results obtained from Raman, IR or neutron scattering and RIM.

Velocities of sound following different directions of propagation have been evaluated from the slopes of acoustic branches. Our results and experimental ultrasonic velocities at 4.2 K are presented in Table 3. The difference between calculated and measured velocities lies within 5% for 7 velocities out of 8. The 9% of discrepancy is obtained for the acoustic branch following the [001] direction. In this direction the longitudinal acoustic branches are non-degenerate although the modes at the **Z** point are degenerate. This behavior has been observed on the two phonon dispersion curves calculated with RIM but seems absent from neutron scattering experiments.

Concerning the phonon modes, the spectrum contains 36 phonons modes at the **G** point as expected from the number of atoms per primitive cell. From group theoretical consideration, irreducible representation for the zone center modes of $C_{4h}$ is given by:

$\Gamma : 3A_g \oplus 3B_g \oplus 3E_g \oplus 5A_u \oplus 3B_u \oplus 5E_u$

$E$ modes are doubly degenerate, $A_g$, $B_g$ and $E_g$ modes are Raman active, $A_u$, and $E_u$ modes are infrared active and $B_u$ is inactive.

The Raman active mode frequencies are presented in Table 4. First of all, it should be noticed this calculation was performed at 0K. The thermal error is not supposed to be very important since, from the results obtained by Raman spectroscopy at 300K and 4.2 K[8], the variation did not exceed 12 cm$^{-1}$. In general the first principle calculated frequencies are lower than experimental results. However, the variation does not exceed 21 cm$^{-1}$ for 10 out of 13 frequencies. The three frequencies showing the highest discrepancy belong to $B_g$ modes with a difference of about 30 cm$^{-1}$. Concerning the first two modes this difference does not induce a change in the classification. But regarding the last $B_g$ mode, we can notice a permutation with the fourth $E_g$ mode. This permutation is visible on phonon dispersion curves obtained by Salaün et al.[11] with RIM and eleven fitting parameters. But, it's absent in the results of Sen et al.[12] using five variable parameters, showing the dependency of results on the number of variable parameters. The vibration modes are presented in Figure 2. The $B_g$ mode calculated at 347 cm$^{-1}$ (Exp.[11] 379 cm$^{-1}$) is mainly due to the translation of the lithium atoms in the direction of the *c axis*. This movement induces a translation of the fluorine atoms along the direction bridging Yttrium and Fluorine atoms. The $E_g$ mode calculated at 356 cm$^{-1}$ (Experimantal value is 373 cm$^{-1}$[11]) is mainly due to the *c axis* rotation of the lithium atoms. These two mode dependent movements agree with the symmetry-adapted motions deduced from group theory analysis[6][7]. We can also notice that the experimental difference between these modes is 6 cm$^{-1}$ which is very close to the error described by Miller et al.[7]. It's interesting to notice that the first $Eg$ phonon mode is difficult to measure experimentally because of polarization effect. This frequency is easily extracted from our results at 149 cm$^{-1}$ (the experimental value is 154 cm$^{-1}$[6][9]).

Concerning the infrared active modes, as LiYF$_4$ is a weak polar crystal, LO/TO mode splitting is experimentally observed. In our calculation LO/TO modes splitting was not taken into account. Consequently, only infrared TO modes were presented in Table 5. From experimental results obtained at 300K and 77K[6], it seems that the effect of the temperature induces a small variation, the disparity did not exceed 6 cm$^{-1}$. In our results obtained at 0 K, four frequencies out of eight are quite close to experimental results obtained by neutron scattering[11]. The difference lies within 3 cm$^{-1}$. For the other modes the most important discrepancy is 25 cm$^{-1}$. Three calculated infrared active modes are higher than experimental one, which is more complicated than the situation of Raman modes mentioned above.

The classification of the phonon symmetry in function of their frequencies is given in Table 6. We can notice only one permutation between two modes below 360 cm$^{-1}$ compared to experimental frequencies, it concerns the inversion of $E_g$ and $A_u$ modes. Above 360 cm$^{-1}$ in addition of the inversion between the $E_g$ and $B_g$ described previously, inversions between Raman and infrared active mode were observed. The first one concerns the intercalation of the $B_g$ between the two last $A_u$ modes compare to experimental results. The second concerns the $E_u$ mode whose frequency is higher than the last $E_g$ and $B_g$. Nevertheless, the agreement between calculated and experimental data is better than RIM calculation.

The phonon density of states $g(w)$ and partial density of state providing the frequency distribution of normal modes are given in Figure 3. As expected from the mass of the different constituents, the density of state can be divided in two parts. *Y* vibrations are dominant between 0 and 250 cm$^{-1}$ and vanish after 417.5 cm$^{-1}$. We can see that in general there is no preferred direction. Exception can be found around 216 cm$^{-1}$ where the movement along *c axis* is privileged. *Li* vibrations begin at 167 cm$^{-1}$ and are dominant between 250 cm$^{-1}$ and 500 cm$^{-1}$. In the range from 250 cm$^{-1}$ to 417.5 cm$^{-1}$, except between 300 cm$^{-1}$ and 350 cm$^{-1}$ where no direction was preferred, the vibrations are along the *c axis* whereas from 417.5cm$^{-1}$ to 500 cm$^{-1}$ movement is in the *ab plan*. We can notice that *F* movements lie between 0 cm$^{-1}$ and 500 cm$^{-1}$ implying that they are first correlated with *Y* and then with *Li*. Movement implying the three different atom types are situated between 167 cm$^{-1}$ and 417.5 cm$^{-1}$. Compare to DOS obtained by RIM method, important difference can be noticed above 250 cm$^{-1}$. The most striking difference concerns the high frequency. Indeed, the DOS calculated in this work stops at 500 cm$^{-1}$ whereas the DOS calculated with RIM stops at 550 cm$^{-1}$ implying *Li* and *F* correlated vibrations.

**IV Conclusion**

This work presents at our knowledge, the first ab initio lattice dynamics calculation of fluoride scheelite. Concerning the phonon dispersion curves, satisfactory agreement with inelastic neutron scattering measurement was obtained. Discrepancies between sound velocities calculated from acoustic branches and ultrasonic measurement do not exceed 300m.s$^{-1}$. Moreover, at the center of the Brillouin zone the error on Raman active modes calculated compared to experimental results does not exceed 9%, the most important error being 33 cm$^{-1}$. One inversion between the last *Bg* mode and the fourth *Eg* mode was put in evidence in comparison with experimental results. Concerning infrared active modes error lies within 8%, the most important error being 25 cm$^{-1}$. Below 360 cm$^{-1}$, only one inversion can be notice compared to experimental results, which is less than in RIM calculations. Important differences between ab initio and RIM calculated DOS were put in evidence mainly above 500 cm$^{-1}$.

**Table 1 : Comparison between calculated at 0K experimental and previous calculated structural parameters of the scheelite LiYF$_4$ structure**

|  | our work | Experimental results | | | Numerical results | | | |
|---|---|---|---|---|---|---|---|---|
|  |  | Ref[32] | Ref[33] | Ref[34] | Ref[35]* | Ref[35]** | Ref[36]*** | Ref[12]**** |
| a,b (Å) | 5.23 | 5.171 | 5.16 | 5.26 | 5.2 | 5.08 | 5.14 | 5.12 |
| c (Å) | 10.82 | 10.748 | 10.74 | 10.94 | 10.85 | 10.54 | 10.82 | 10.67 |
| c/a | 2.07 | 2.08 | 2.08 | 2.08 | 2.09 | 2.07 | 2.11 | 2.08 |
| Density (Mg.m$^{-3}$) | 3.86 | 3.98 | 4.00 | 3.78 | 3.89 | 4.20 | 4.00 | 4.09 |

*: DFT/PAW/GGA
**: DFT/PAW/LDA
***: DFT/OLCAO/LDA
****: RIM

**Table 2 : Comparison between experimental and calculated interatomic distances**

| Ion pair | Distance (Å) | | | |
|---|---|---|---|---|
| | our Work | Experimental results | | |
| | | Ref[32] | Ref[33] | Ref[34] |
| Li-F | 1.92 | 1.89 | 1.89 | 1.73 |
| Y-F | 2.26 | 2.24 | 2.24 | 2.4 |
| Y-F | 2.32 | 2.29 | 2.29 | 2.45 |
| F-F | 2.62 | 2.60 | 2.59 | 2.71 |
| F-F | 2.78 | 2.75 | 2.74 | 2.83 |
| F-F | 2.78 | 2.76 | 2.76 | 2.84 |
| F-F | 2.86 | 2.82 | 2.83 | 3.03 |
| Li-F | 2.93 | 2.9 | 2.89 | 2.96 |
| F-F | 2.98 | 2.95 | 2.95 | 3.06 |

**Table 3 : Sound velocities in characteristic direction of propagation vector in reciprocal lattice.**

| Direction of propagation vector in reciprocal lattice | Branches | Sound velocity Exp.[37] | Sound velocity Calc. |
|---|---|---|---|
| [001] | Longitudinal | 6346 | 6579 |
|  | Transversal | 3276 | 2976 |
| [100] | Longitudinal | 3276 | 3109 |
|  | Transversal | 5576 | 5455 |
|  | Transversal | - | 2053 |
| [110] | Longitudinal | 3276 | 3173 |
|  | Transversal | 5345 | 5146 |
|  | Transversal | 2727 | 2814 |

**Table 4 : Frequencies of Raman active phonons of scheelite LiYF$_4$.**

| Phonon symmetry | Frequencies (cm$^{-1}$) | | | | | | | | |
|---|---|---|---|---|---|---|---|---|---|
| | Our work | Experimental results | | | | | | Numerical results[†] | |
| | | ref[11]* | ref[9]** | ref[6]** | ref[8]** | ref[7]** | ref[10]** | ref[11] | ref[12] |
| Ag | 145 | 151 | 151 | 151 | - | - | - | 149 | 151 |
| | 244 | 265 | 265 | 265 | 269 | 264 | 261 | 239 | 250 |
| | 416 | 427 | 426 | 427 | 426 | 425 | 422 | 419 | 442 |
| Bg | 170 | 174 | 173 | 174 | 177 | 177 | 167 | 188 | 174 |
| | 213 | 246 | 246 | 246 | 251 | 248 | 241 | 258 | 268 |
| | 314 | 327 | 326 | 327 | 331 | 329 | 323 | 319 | 297 |
| | 347 | 379 | 379 | 379 | 382 | 382 | 375 | 369 | 391 |
| | 398 | 427 | 426 | 427 | 430 | 427 | 423 | 423 | 428 |
| Eg | 149 | 154 | 154 | 154 | 158 | 153 | 149 | 152 | 153 |
| | 188 | 199 | 198 | 199 | 203 | 199 | 193 | 212 | 207 |
| | 306 | 326 | 326 | 326 | 329 | 329 | 322 | 312 | 318 |
| | 356 | 373 | 372 | 373 | 376 | 368 | 369 | 378 | 380 |
| | 448 | 447 | 446 | 447 | 450 | 446 | 442 | 448 | 449 |

[†]: RIM model
*: inelastic neutron scattering
**: Raman

**Table 5 : Frequencies of infrared active phonons of scheelite LiYF$_4$.**

| Phonon symmetry | Frequencies (cm-1) | | | | | | |
|---|---|---|---|---|---|---|---|
| | Our work | Experimental results | | | | Numerical results[†] | |
| | | ref[11]* | ref[6]** | ref[7]*** | ref[8]**** | ref[11] | ref[12] |
| Au (TO) | 194 | 196 | 198 | 195 | 192 | 192 | 199 |
| | 231 | 251 | 251 | 252 | 255 | 272 | 251 |
| | 342 | 341 | 339 | 396 | 404 | 324 | 316 |
| | 351 | 370 | 372 | 490 | 489 | 403 | 403 |
| Eu (TO) | 140 | 137 | 137 | 143 | 144 | 156 | 141 |
| | 274 | 294 | 294 | 293 | 262 | 279 | 263 |
| | 300 | 325 | 323 | 326 | 314 | 322 | 319 |
| | 419 | 418 | 418 | 424 | 413 | 424 | 410 |

[†]: RIM model
*: inelastic neutron scattering 300 K
**: reflexion IR (300K)
***: adsorption IR (77K)
****: adsorption IR (330K)

**Table 6: Classification of the phonon symmetry in function of their frequencies**

| Our Work | Experimental results | | | | Numerical results[†] | |
|---|---|---|---|---|---|---|
| | ref[11] | ref[6] | ref[7] | ref[8] | ref[11] | ref[12] |
| **Eu(I)** | **Eu(I)** | **Eu(I)** | **Eu(I)** | **Eu(I)** | Ag(R) | **Eu(I)** |
| Ag(R) | Ag(R) | Ag(R) | Ag(R) | Ag(R) | Eg(R) | Ag(R) |
| Eg(R) | Eg(R) | Eg(R) | Eg(R) | Eg(R) | **Eu(I)** | Eg(R) |
| Bg(R) | Bg(R) | Bg(R) | Bg(R) | Bg(R) | Bg(R) | Bg(R) |
| Eg(R) | **Au(I)** | **Au(I)** | **Au(I)** | **Au(I)** | **Au(I)** | **Au(I)** |
| **Au(I)** | Eg(R) | Eg(R) | Eg(R) | Eg(R) | Eg(R) | Eg(R) |
| Bg(R) | Bg(R) | Bg(R) | Bg(R) | Bg(R) | Ag(R) | Ag(R) |
| **Au(I)** | **Au(I)** | **Au(I)** | **Au(I)** | **Au(I)** | Bg(R) | **Au(I)** |
| Ag(R) | Ag(R) | Ag(R) | Ag(R) | **Eu(I)** | **Au(I)** | **Eu(I)** |
| **Eu(I)** | **Eu(I)** | **Eu(I)** | **Eu(I)** | Ag(R) | **Eu(I)** | Bg(R) |
| **Eu(I)** | **Eu(I)** | **Eu(I)** | **Eu(I)** | **Eu(I)** | Eg(R) | Bg(R) |
| Eg(R) | Eg(R) | Eg(R) | Eg(R) | Eg(R) | Bg(R) | **Au(I)** |
| Bg(R) | Bg(R) | Bg(R) | Bg(R) | Bg(R) | **Eu(I)** | Eg(R) |
| **Au(I)** | **Au(I)** | **Au(I)** | Eg(R) | Eg(R) | **Au(I)** | **Eu(I)** |
| Bg(R) | **Au(I)** | **Au(I)** | Bg(R) | Bg(R) | Bg(R) | Eg(R) |
| **Au(I)** | Eg(R) | Eg(R) | **Au(I)** | **Au(I)** | Eg(R) | Bg(R) |
| Eg(R) | Bg(R) | Bg(R) | **Eu(I)** | **Eu(I)** | **Au(I)** | **Au(I)** |
| Bg(R) | **Eu(I)** | **Eu(I)** | Bg(R) | **Au(I)** | Bg(R) | **Eu(I)** |
| Ag(R) | Bg(R) | Bg(R) | Ag(R) | Bg(R) | Ag(R) | Bg(R) |
| **Eu(I)** | Ag(R) | Ag(R) | Eg(R) | Ag(R) | **Eu(I)** | Ag(R) |
| Eg(R) | Eg(R) | Eg(R) | **Au(I)** | Eg(R) | Eg(R) | Eg(R) |

**Figure caption:**

**Figure 1 :** Calculated phonon dispersion curves of scheelite LiYF$_4$. The path is defined in direction of the quadratic Brillouin zone. The labels are adapted to the symmetry of the active structure. Z (1/2 1/2 –1/2), G=$\Gamma$ (0 0 0), X (0 0 1/2), P(1/4 1/4 1/4), N(0 0.5 0)

**Figure 2 :** Schematics of corresponding atomic displacements for the Bg and Eg Raman active modes respectively at 347 cm$^{-1}$ and 356 cm$^{-1}$. Y atoms are black balls, Li atoms are dark grey and F atoms are light grey.

**Figure 3 :** Total and partial phonon density of state calculated for Y, Li and F atoms and x ,y, z Cartesian directions.

**Figure 1**

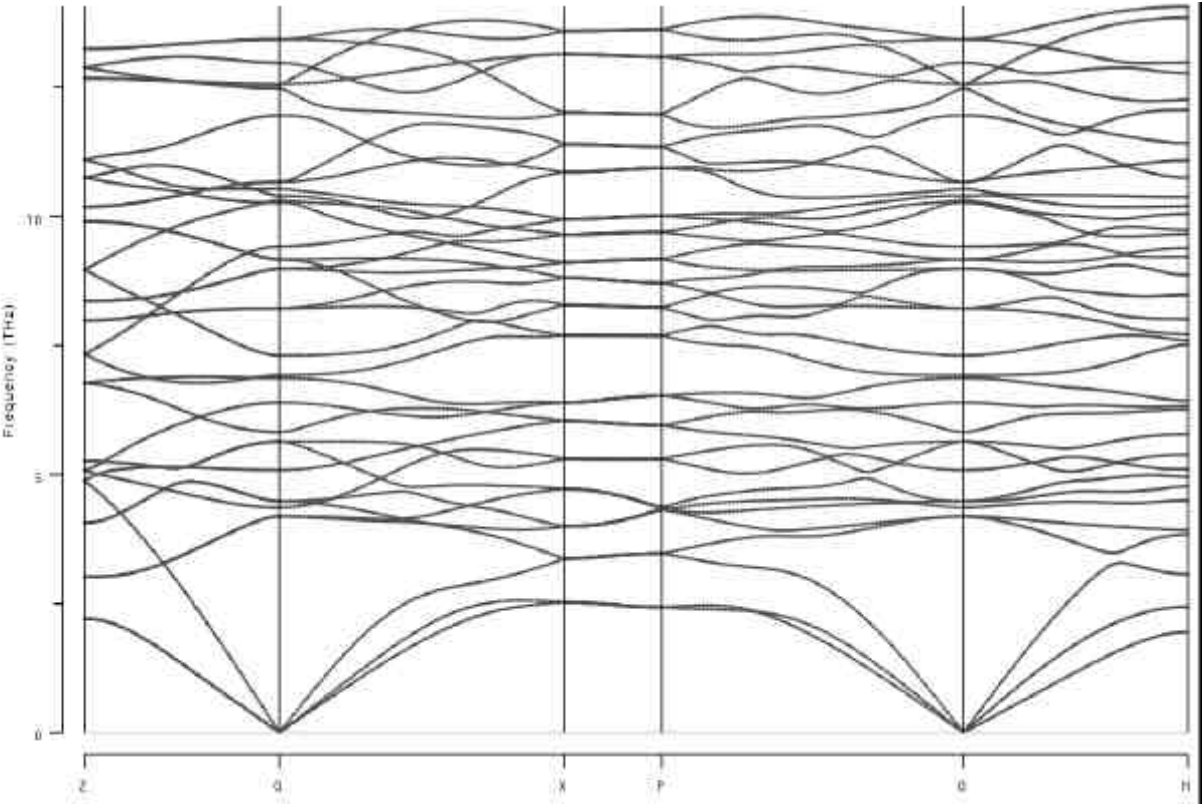

Figure 2

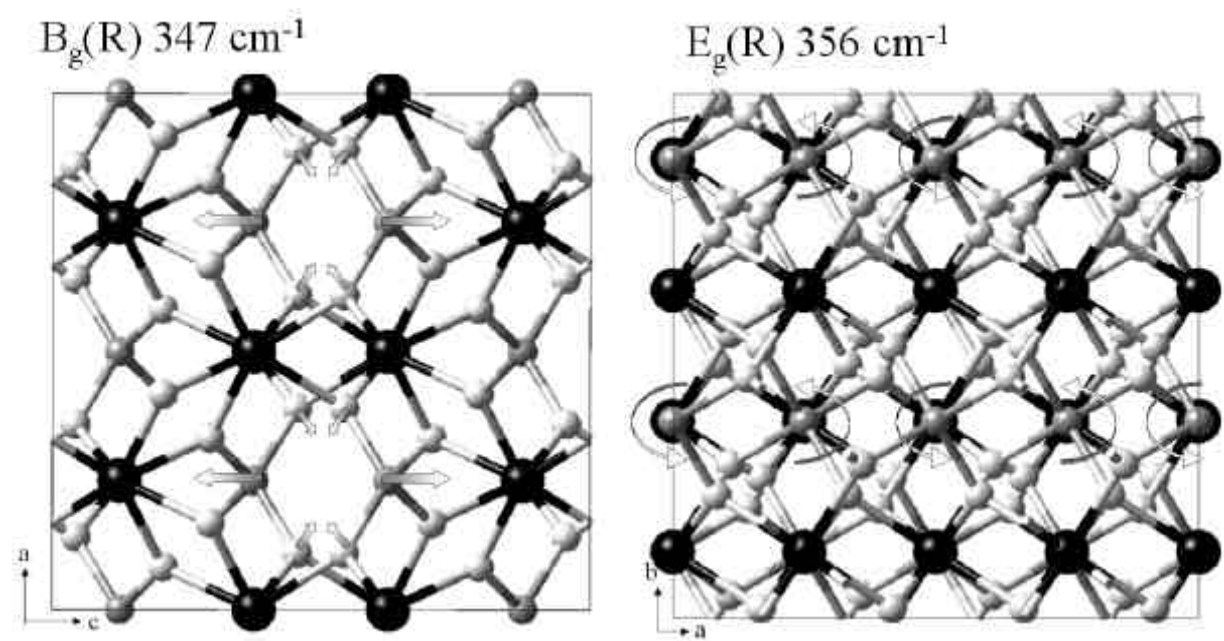

Figure 3

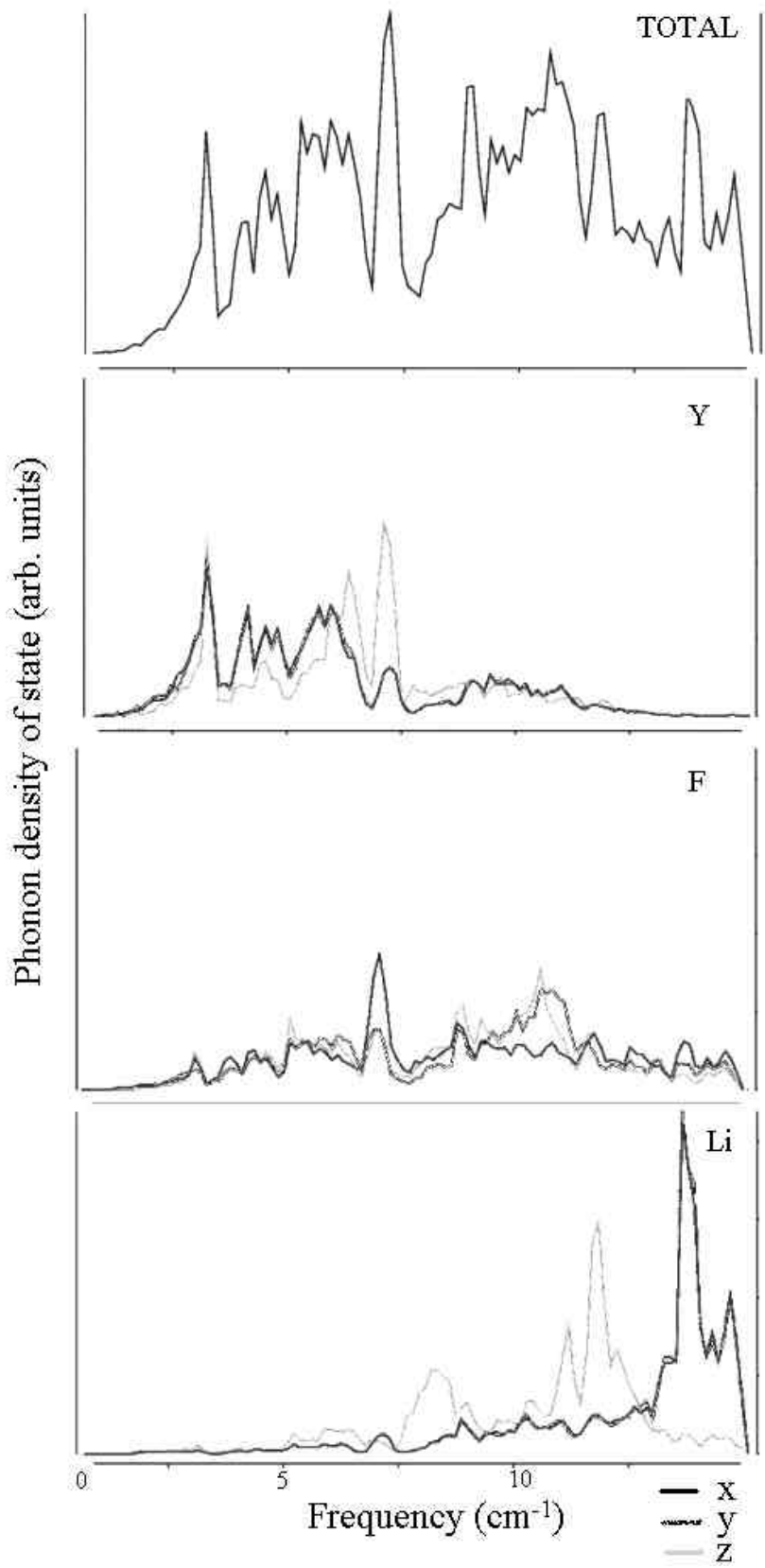